\documentclass[9pt,conference]{IEEEtran}

\usepackage[preprint]{waspaa25}

\usepackage{bm} 

\title{Two Views, One Truth: Spectral and Self-Supervised Features Fusion
for Robust Speech Deepfake Detection}

\name{Yassine El Kheir$^{1,2}$,
      Arnab Das$^{1,3}$,
      Enes Erdem Erdogan$^{1}$,
      Fabian Ritter-Guttierez$^{4}$,
      Tim Polzehl$^{1,3}$,
      Sebastian Möller$^{1,2}$}

\address{$^{1}$Speech and Language Technology, DFKI, Germany\\
$^{2}$Quality and Usability Lab, Technical University of Berlin, Germany\\
$^{3}$AI Team, Gretchen AI, Germany\\
$^{4}$Nanyang Technological University, Singapore
}

\begin{document}

\maketitle

\begin{abstract}
Recent advances in synthetic speech have made audio deepfakes increasingly realistic, posing significant security risks. Existing detection methods that rely on a single modality, either raw waveform embeddings or spectral‐based features, are vulnerable to non‐spoof disturbances and often overfit to known forgery algorithms, resulting in poor generalization to unseen attacks. To address these shortcomings, we investigate hybrid fusion frameworks that integrate self‐supervised learning (SSL)-based representations with handcrafted spectral descriptors (e.g., MFCC, LFCC, CQCC). By aligning and combining complementary information across modalities, these fusion approaches capture subtle artifacts that single‐feature approaches typically overlook. We explore several fusion strategies, including simple concatenation, cross‐attention, mutual cross‐attention, and a learnable gating mechanism, to optimally blend SSL features with fine‐grained spectral cues. We evaluate our approach on four challenging public benchmarks (LA19, DF21, ITW, ASV5) and report generalization performance. All fusion variants consistently outperform an SSL‐only baseline, with the cross‐attention strategy achieving the best generalization with a 38\% relative reduction in equal error rate (EER). These results confirm that joint modeling of waveform and spectral views produces robust, domain‐agnostic representations for audio deepfake detection. 
\end{abstract}



\section{Introduction}


Recent progress in generative artificial intelligence has significantly transformed the field of ultra-realistic speech synthesis.
Advances in state-of-the-art text-to-speech (TTS) and voice conversion (VC) technologies have made it feasible to produce speech outputs that closely mimic natural human utterances \cite{pham2025comprehensive}. 
Moreover, the emergence of large language models (LLMs) into audio and speech generation \cite{hao2025boosting, zhang2023speechgpt, rubenstein2023audiopalm} workflows has further elevated the potential of generating complex speech signals with high fidelity and nuance.
Despite these encouraging advancements, they have also introduced significant challenges. 
The increasing accessibility of tools capable of generating highly realistic speech has raised concerns about their misuse by malicious actors. 
These technologies are being exploited to disseminate misinformation, incite hate speech, and even support acts of terrorism activities \cite{meaker2023deepfake} that pose a direct threat to societal trust and the integrity of public discourse. 
Furthermore, synthetic speech has become a tool for financial fraud; notably, a recent report highlighted that voice cloning technologies have impacted approximately 7.7\% of individuals, including high-profile cases such as a CEO in the United Kingdom \cite{mcafee2023beware}.
Hence, the need for robust countermeasure systems that help humans discern natural and spoofed speech is paramount.

Nonetheless, detecting synthetic or spoofed speech remains a significant challenge. 
One major obstacle is the rapid evolution of speech synthesis techniques, which often renders existing detection methods obsolete in a short time.
Additionally, the wide variability in generated speech, arising from differences in synthesis tools, transmission channels, audio codecs, and environmental noise, further undermines the effectiveness of current models.
As a result, the pursuit of detection methods that are not only highly accurate but also generalizable and robust across diverse conditions continues to be a critical research priority.

\begin{figure}
    \centering
    \includegraphics[width=\linewidth]{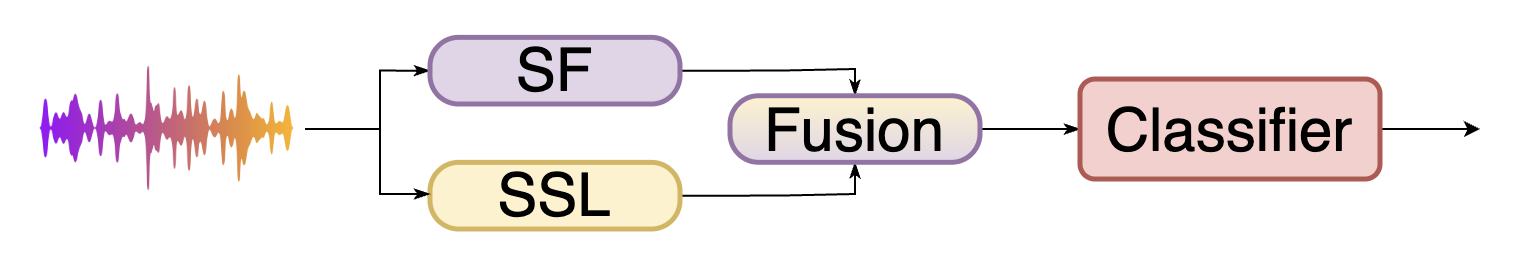}
    \caption{Overall framework of our approach. SF stands for Spectral Features, and SSL for Self-Supervised Features. }
    \label{fig:model}
\end{figure}

Early approaches to synthetic speech detection primarily relied on manually engineered spectral features such as Linear Frequency Cepstral Coefficients (LFCC) \cite{chen2021ur}, Mel-Frequency Cepstral Coefficients (MFCC) \cite{caceres2021biometric}, Constant Q Cepstral Coefficients (CQCC) \cite{das2021known}, Constant-Q Transform (CQT) \cite{tomilov2021stc}, Short-Time Fourier Transform (STFT) \cite{muller2022does}, and Mel-spectrograms \cite{alenin2022subnetwork}. 
These features capture high-frequency artifacts that are typically fed into classifiers, often convolutional neural networks (CNNs) or multilayer perceptrons (MLPs), to perform binary classification between genuine (bonafide) and spoofed speech. 
A variety of CNN-based architectures have been explored for this task, including ResNet \cite{chen2021pindrop}, Inception \cite{muller2022does}, Res2Net \cite{dong2023multi}, ECAPA-TDNN \cite{wang2021comparative}, LCNN \cite{wang2024multi}.

In parallel, end-to-end approaches operate directly on raw audio waveforms \cite{caceres2021biometric, ge2021raw}. In particular, the AASIST \cite{jung2022aasist} model processes raw speech using fixed sinc-convolutional filter banks, followed by a spectro-temporal graph neural network (GNN) equipped with attention mechanisms, demonstrating improved performance over earlier models. 

\begin{figure*}
    \centering
    \includegraphics[width=\linewidth]{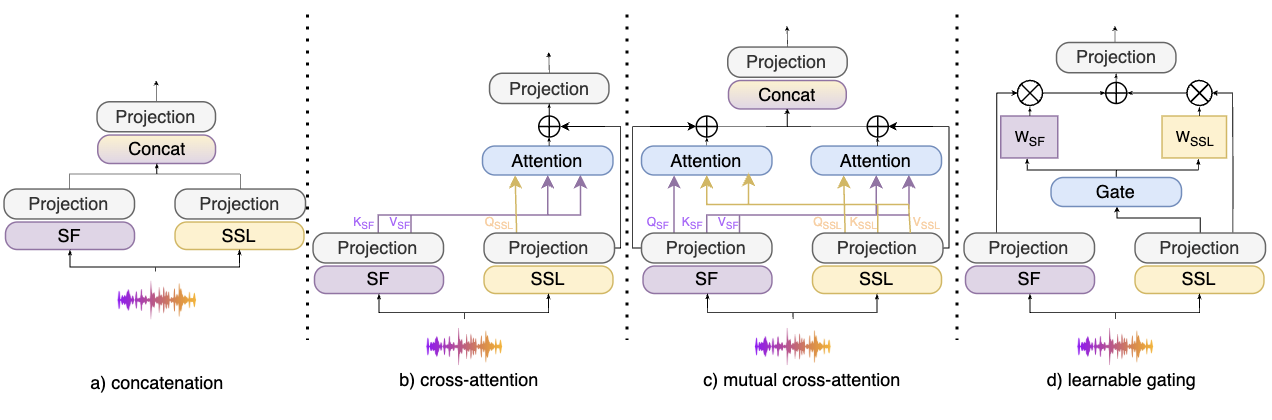}
    \vspace{-0.7cm}
    \caption{Illustration of the different investigated fusion strategies, combining spectral features with self-supervised learning representations.}
    \label{fig:fusions}
\end{figure*}

More recent developments leverage self-supervised learning (SSL) techniques to obtain robust audio representations using transformer-based encoders such as Wav2vec2.0 \cite{xie2023learning}, XLSR \cite{wang2024can}, WavLM \cite{yang2024robust}, HuBERT \cite{yang2024robust}. These SSL-based methods have shown substantial improvements over traditional approaches based on handcrafted features, yet suffer as well from a lack of generalization \cite{muller2024harder}. 

Hybrid models that integrate both SSL representations and traditional handcrafted spectral features have demonstrated efficacy in various tasks, including ASR \cite{berrebbi2022combining} and speaker verification \cite{peng2024fine}, as well as recently in spoofing detection \cite{jin2025wave}. Hand-crafted features such as MFCCs, which offer data-independence and task-robust features and are sensitive to synthetic artifacts, can enhance overall system robustness when combined with SSL features.


However, key questions remain open regarding which combinations of features are most effective for deepfake speech detection, the optimal mechanisms for integrating handcrafted and SSL-based representations, and the relative contribution or weighting of each feature stream in a trained model, highlighting important directions for future research.

In line with these research gaps, this paper makes the following key contributions:
\begin{itemize}
    \item We investigate the effectiveness of three widely used spectral features—MFCC, LFCC, and CFCC—when combined with self-supervised speech representations learned via Wav2Vec2.0 XLSR-53.
    \item We explore various fusion strategies, including concatenation, attention-based, and gating mechanisms, to identify optimal approaches for integrating spectral and SSL-derived features.
    \item We analyze the relative contribution of each feature stream (spectral vs. SSL) during inference to understand their impact within the trained model.
\end{itemize}
The proposed methodology is detailed in Section~\ref{sec:method}, followed by the experimental setup in Section~\ref{sec:exp}, and the corresponding results and analysis in Section~\ref{sec:results}.

\section{Methodology}
\label{sec:method}

The overall architecture of our proposed system is shown in Figure \ref{fig:model}. The input audio waveform is fed into two parallel streams that extract complementary speech representations: one based on SSL model and the other on hand-crafted spectral features (SF). These representations are fused before being passed to a graph neural network-based classifier, AASIST \cite{jung2022aasist}.

\subsection{Speech Representations}

\paragraph*{\textbf{Spectral Features (SF)}} 
In this work, we focus on the three most widely used hand-crafted feature extractors in deepfake detection \cite{zhang2025audio}: Mel-Frequency Cepstral Coefficients (MFCC), Linear-Frequency Cepstral Coefficients (LFCC), and Constant-Q Cepstral Coefficients (CQCC). All three apply cepstral analysis to extract short-term spectral features but differ in their frequency scaling. MFCCs are cepstral features that use a Mel scale to emphasize low-frequency components. LFCCs adopt a linear frequency scale, preserving high-frequency detail and enhancing sensitivity to spectral artifacts, particularly in unseen attack scenarios. Lastly, CQCCs, derived from the constant-Q transform, combine multiresolution time–frequency analysis with cepstral representation, making them effective for capturing complex spectral structures.

\paragraph*{\textbf{Self-Supervised Learning Features (SSL)}}
For the SSL branch, we employ Wav2Vec2.0 XLSR-53, a transformer-based model trained on large-scale, unlabeled speech corpora using contrastive objectives. Wav2Vec2.0 XLSR-53 demonstrates strong  capabilities in speech deepfake detection \cite{kheir2025comprehensive}. These representations often capture rich contextual and prosodic information, which are essential for identifying subtle artifacts introduced by generative speech models.

\subsection{Fusion Strategies}
\label{sec:fusion}
We extract two parallel feature streams for a given segment \textbf{$S$}: (i) \textbf{SSL features} from pre-trained Wav2Vec2.0 model, denoted \( \mathbf{f_{\text{SSL}}(S)} \in \mathbb{R}^{T_{\text{SSL}} \times D_{\text{SSL}}} \); and (ii) \textbf{Spectral Features (SF)} that can be MFCC, LFCC, or CQCC, denoted \( \mathbf{f_{\text{SF}}(S)} \in \mathbb{R}^{T_{\text{SF}} \times D_{\text{SF}}} \). Here, \( T_{\text{SSL}}, T_{\text{SF}} \) are the number of time frames, and \( D_{\text{SSL}}, D_{\text{SF}} \) are their respective feature dimensions.

To fuse these representations, we first align them to a common temporal and feature resolution. We set a common frame count \( T = T_{\text{SSL}} \) and feature dimension \( D \) by resampling and projecting. Specifically, we downsample the SF stream such that \( T_{\text{SF}} \rightarrow T \), and apply a linear projection \( \in \mathbb{R}^{D_{\text{SSL}} \times D} \) and \( \in \mathbb{R}^{D_{\text{SF}} \times D} \) to map SSL and SF features into \( \mathbb{R}^{T \times D} \). After this alignment, we have:

\[
f_{\text{SF}}(S),\,f_{\text{SSL}}(S) \in \mathbb{R}^{T \times D}
\]

Let \( \mathbf{f}_{\text{SF}}, \mathbf{f}_{\text{SSL}} \in \mathbb{R}^{T \times D} \) denote the aligned feature matrices. We explore four fusion strategies as shown in Figure \ref{fig:fusions}:

\paragraph*{\textbf{Concatenation}} We concatenate the features along the feature dimension and project them back into \( \mathbb{R}^{T \times D} \):
\[
\mathbf{h}_{\text{concat}} = \text{Linear}([\mathbf{f}_{\text{SF}}; \mathbf{f}_{\text{SSL}}])
\]
where \([\,;\,]\) denotes concatenation along the feature dimension.

\paragraph*{\textbf{Cross-Attention (SSL \textrightarrow SF)}} SSL features attend to SF features. Define projections:
\[
\mathbf{Q}_{\text{SSL}} = \mathbf{f}_{\text{SSL}} W_Q,\quad \mathbf{K}_{\text{SF}} = \mathbf{f}_{\text{SF}} W_K,\quad \mathbf{V}_{\text{SF}} = \mathbf{f}_{\text{SF}} W_V
\]
Then,
\[
\mathbf{H}_{\text{SSL} \rightarrow \text{SF}} = \text{Softmax}\left( \frac{\mathbf{Q}_{\text{SSL}} \mathbf{K}_{\text{SF}}^\top}{\sqrt{D}} \right) \mathbf{V}_{\text{SF}} + \mathbf{f}_{\text{SSL}}
\]

\paragraph*{\textbf{Mutual Cross-Attention}} We compute attention in both directions:
\[
\mathbf{Q}_{\text{SF}} = \mathbf{f}_{\text{SF}} W_Q,\quad \mathbf{K}_{\text{SSL}} = \mathbf{f}_{\text{SSL}} W_K,\quad \mathbf{V}_{\text{SSL}} = \mathbf{f}_{\text{SSL}} W_V
\]
\[
\mathbf{H}_{\text{SF} \rightarrow \text{SSL}} = \text{Softmax}\left( \frac{\mathbf{Q}_{\text{SF}} \mathbf{K}_{\text{SSL}}^\top}{\sqrt{D}} \right) \mathbf{V}_{\text{SSL}} + \mathbf{f}_{\text{SF}}
\]
\[
\mathbf{h}_{\text{mutual}} = \text{Linear}([\mathbf{H}_{\text{SF} \rightarrow \text{SSL}};\, \mathbf{H}_{\text{SSL} \rightarrow \text{SF}}])
\]

\paragraph*{\textbf{Learnable Gating}} SF and SSL are treated as two experts and we learn per-frame weights:
\[
\mathbf{w}(S) = \text{Softmax}(\mathbf{f}_{\text{SSL}} W_G),\quad \mathbf{w}(S) \in \mathbb{R}^{T \times 2}
\]
Let \( \mathbf{w}_{\text{SF}}, \mathbf{w}_{\text{SSL}} \in \mathbb{R}^{T} \) be the two columns of \( \mathbf{w}(S) \). Then the fused feature is:
\[
f_{\text{fuse}}(S)_t = w_{\text{SF}}(t)\,f_{\text{SF}}(S)_t + w_{\text{SSL}}(t)\,f_{\text{SSL}}(S)_t
\]
or in matrix form:
\[
\mathbf{h}_{\text{fuse}} = \sum_{i \in \{\text{SF}, \text{SSL}\}} \mathbf{w}_i \odot \mathbf{f}_i
\]
where \( \odot \) denotes element-wise multiplication applied per frame. Since the weights are normalized per frame, this fusion is both adaptive and interpretable. Each fused representation \( \mathbf{h}_{\text{fuse}} \) is passed to a classifier. The gating weights can be visualized to interpret streams fusions over time.

\subsection{Classifier: AASIST}
AASIST~\cite{jung2022aasist} is a SOTA audio anti-spoofing back-end classifier that integrates spectral and temporal information using heterogeneous graph-based attention and pooling mechanisms. Its architecture comprises a Graph Attention Layer (GAT) for computing attention over spectral and temporal features, a Heterogeneous Graph Attention Layer (HtrgGAT) to refine these features, a graph pooling layer for selecting salient nodes, residual blocks with convolutional layers and SELU activations, and an attention mechanism to extract meaningful representations from the encoded data. This design enables AASIST to effectively capture complex patterns associated with spoofing attacks across diverse audio domains.





\section{Experimental Setup}
\label{sec:exp}




    

\subsection{Datasets and Evaluation Metrics}

To assess the effectiveness and generalization capability of our proposed approach, we conduct experiments on four benchmark datasets: ASVspoof LA19 (LA19)~\cite{wang2020asvspoof}, ASVspoof DF21 (DF21)~\cite{yamagishi2021asvspoof}, In-The-Wild (ITW)~\cite{muller2022does}, and ASVspoof 5 (ASV5)~\cite{wang2024asvspoof5}. Our models are primarily trained on the LA19 training partition, a well-established benchmark in the speech anti-spoofing community. We then evaluate the trained models across all four datasets to test their robustness and cross-domain generalization.

\begin{itemize}
    \item \textbf{ASVspoof DF21 (DF21)}: This dataset introduces a diverse set of spoofing attacks generated using deepfake synthesis techniques and encoded with lossy compression codecs. It reflects real-world scenarios where audio is transmitted or stored under bandwidth constraints. The dataset includes 14,869 bona fide utterances and 519,059 spoofed samples.

    \item \textbf{In-The-Wild (ITW)}: ITW features a realistic distribution of spoofed and genuine speech collected from publicly available online sources such as podcasts and political speeches. The dataset comprises 17.2 hours of fake and 20.7 hours of authentic audio, amounting to 31,779 utterances with an average duration of 4.3 seconds. All recordings involve English-speaking celebrities and politicians.

    \item \textbf{ASVspoof 5 (ASV5)}: The ASV5 corpus represents the most recent and challenging benchmark in the domain, incorporating 32 advanced spoofing attack types, including adversarially crafted samples. It contains 138,688 bonafide and 542,086 spoofed utterances, designed to evaluate the limits of current anti-spoofing systems.

\end{itemize}

We evaluate model performance using the Equal Error Rate (EER), the standard metric in anti-spoofing tasks.


\subsection{Implementation Details}
\vspace{-0.1cm}
In the pre-processing stage, all audio samples are first pre-emphasized using a coefficient of 0.97. Each sample is then either truncated or zero-padded to a fixed length of approximately 4 seconds (64,600 samples) to ensure uniformity across the dataset. No voice activity detection or amplitude normalization is applied.

For the Wav2Vec2.0 model, we initialize parameters using the official pre-trained weights\footnote{https://github.com/facebookresearch/fairseq/blob/main/examples/wav2vec}, so the $f_{\text{SSL}}$ has a shape of \(201 \times 1024\). For spectral features, we use a 25-ms Hamming window with a 10-ms frame shift. For each frame, we compute MFCC using a 20-filter Mel-scale bank, LFCC using a 20-filter linear-scale bank, and CQCC derived from a Constant-Q transform with 96 bins per octave. We then append first- and second-order derivatives to each 20-dimensional vector, producing a 60-dimensional feature vector per frame and a final tensor $f_{\text{SF}}$ of shape \(402 \times 60\). The projection dimension $D$ is set to 128.

Model training is performed using the Adam optimizer~\cite{diederik2014adam} with $\beta_1 = 0.9$ and $\beta_2 = 0.999$. A step-based learning rate scheduler is used to facilitate faster convergence. All models are trained for 50 epochs with a batch size of 32, an initial learning rate of \(1 \times 10^{-6}\), and a weight decay of \(1 \times 10^{-4}\). The cross-entropy loss is used as the objective function.

All training runs are executed on a single NVIDIA H100 GPU. To ensure robustness and account for variability, each experiment is repeated three times using different random seeds.

\section{Results and Analysis}
\label{sec:results}
\begin{table}[]
\caption{EER\% across four evaluation datasets (LA19, DF21, ITW, ASV5) for different fusion strategies between spectral features (MFCC, LFCC, CQCC) and SSL representations. “\textbf{No Fusion}” only spectral features are used. The \textbf{AVG} column reports the average performance across all datasets. Values in bold indicate the best performance on each dataset.}
\label{tab:table1}
\scalebox{1.2}{
\centering
\begin{tabular}{lrrrrr}
\toprule

                  & \multicolumn{1}{c}{\textbf{LA19}} & \multicolumn{1}{c}{\textbf{DF21}} & \multicolumn{1}{c}{\textbf{ITW}} & \multicolumn{1}{c}{\textbf{ASV5}} & \multicolumn{1}{c}{\textbf{AVG}} \\ \hline
\textbf{Baseline} & 0.28                              & 5.29                              & 14.03                            & 23.88                             & 10.87                            \\ \hline

                  & \multicolumn{5}{c}{\textit{\textbf{No Fusion}}} \\ \hline
\textbf{MFCC}     & 13.35                            & \textbf{33.15}                             & 39.20                 & \textbf{42.95}                 & 32.16                          \\ \cline{2-6} 
\textbf{LFCC}     & 5.62                             & 34.28                              & 38.46                            & 44.02                             & 30.59                            \\ \cline{2-6} 
\textbf{CQCC}     & \textbf{4.63}                              & 34.62                   & \textbf{33.68}                             & 44.51                             & \textbf{29.36}                 \\ \hline

                  & \multicolumn{5}{c}{\textit{\textbf{Concatenation}}}                                                                                                                                      \\ \hline
\textbf{MFCC}     & 0.42                              & 4.68                              & \textbf{8.48}                    & \textbf{18.93}                    & 8.13                             \\ \cline{2-6} 
\textbf{LFCC}     & 0.82                              & 4.19                              & 9.35                             & 20.06                             & 8.61                             \\ \cline{2-6} 
\textbf{CQCC}     & 0.42                              & \textbf{4.09}                     & 8.62                             & 19.24                             & \textbf{8.09}                    \\ \hline
                  & \multicolumn{5}{c}{\textit{\textbf{Cross-Attention (SSL \textrightarrow SF)}}}                                                                                                                                          \\ \hline
\textbf{MFCC}     & 0.46                              & 3.55                              & 7.11                             & 21.25                             & 8.09                             \\ \cline{2-6} 
\textbf{LFCC}     & \textbf{0.28}                     & 3.15                              & 6.87                             & 20.85                             & 7.79                             \\ \cline{2-6} 
\textbf{CQCC}     & 0.40                              & \textbf{2.71}                     & \textbf{6.03}                    & \textbf{18.08}                    & \textbf{6.80}                    \\ \hline
                  & \multicolumn{5}{c}{\textit{\textbf{Mutual Cross-Attention}}}                                                                                                                                         \\ \hline
\textbf{MFCC}     & 0.36                              & 3.49                              & 7.89                             & 20.89                             & 8.16                             \\ \cline{2-6} 
\textbf{LFCC}     & 0.58                              & 3.53                              & 8.40                             & \textbf{18.68}                    & 7.80                             \\ \cline{2-6} 
\textbf{CQCC}     & \textbf{0.26}                     & \textbf{3.09}                     & \textbf{7.04}                    & 18.75                             & \textbf{7.29}                    \\ \hline
                  & \multicolumn{5}{c}{\textit{\textbf{Learnable Gating}}}                                                                                                                                 \\ \hline
\textbf{MFCC}     & 0.61                              & \textbf{3.39}                     & 8.75                             & 19.46                             & \textbf{8.05}                             \\ \cline{2-6} 
\textbf{LFCC}     & \textbf{0.39}                     & 3.77                              & \textbf{7.88}                    & 21.90                             & 8.49                             \\ \cline{2-6} 
\textbf{CQCC}     & 0.53                              & 3.70                              & 9.24                             & \textbf{20.00}                    & 8.37                             \\ \bottomrule
\end{tabular}}
\end{table}

\subsection{Evaluating Fusion Approaches}

\noindent As shown in Table~\ref{tab:table1}, under the “\textbf{No Fusion}” setting, models relying solely on handcrafted spectral features (e.g., MFCC, LFCC, CQCC) perform poorly in both controlled and in-the-wild conditions, with average EERs no lower than 29.36\%. By contrast, the baseline system, Wav2Vec2.0 XLSR-53 + AASIST \cite{laakkonen2025generalizable}, reduces the average EER to 10.87\%, corroborating previous findings that SSL-based features alone outperform spectral features alone. Nevertheless, this wave‐only baseline remains susceptible to domain shifts, exhibiting an EER increase of 13.55\% on the ITW benchmark and 23.40\% on ASV5.

\noindent On the other hand, all examined fusion approaches, concatenation, cross‐attention (SSL→SF), mutual cross‐attention, and learnable gating, exploit additional spectral cues to outperform the wave‐only baseline across every evaluation dataset, and with different proposed spectral features. They achieve relative improvements ranging from 11.5\% to 48.77\% on DF21, 33.35\% to 57.00\% on ITW, and 8.2\% TO 24.28\% on ASV5. These findings show that combining the two features captures complementary information beneficial for deepfake detection.

Among these, \textbf{Cross-Attention (SSL → SF)} with CQCC yields the best overall performance, achieving the lowest average EER of 6.80\% and outperforming all fusion systems on DF21 and ITW with EERs of 2.71\% and 6.03\%, respectively. This strong performance can be attributed to the ability of cross-attention to dynamically learn which components of the spectral features are most relevant when conditioned on the richer SSL representations. By attending selectively to complementary cues in the handcrafted features, the model effectively enhances discriminative capacity in challenging scenarios. Mutual cross-attention also shows strong performance, with CQCC-based fusion reaching an average EER of 7.29\%. Simple concatenation and learnable gating yield noticeable gains but fall short of the attention-based methods, which better capture cross-modal dependencies.


Across all fusion schemes, on average across datasets, CQCC consistently outperforms MFCC and LFCC, demonstrating its effectiveness in modeling the fine-grained spectral patterns essential for robust spoofing detection, which aligns with previous findings \cite{todisco2017constant}.

\subsection{Analysis of Learnable Gating Weights}
\label{sec:moe_analysis}

\begin{figure}
    \centering
    \includegraphics[width=0.7\linewidth]{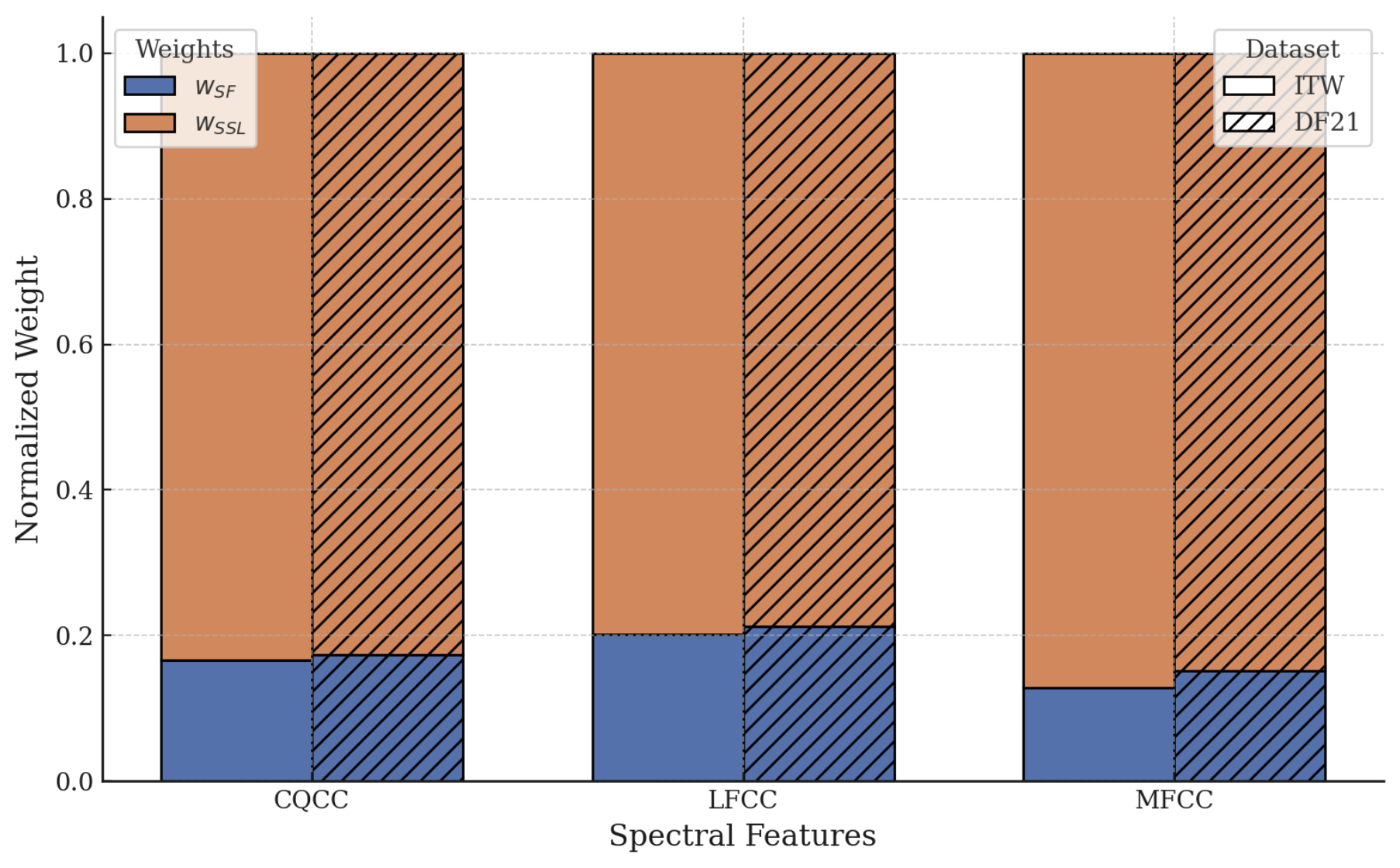}
    \caption{Normalized learnable gating weights for spectral features and SSL across ITW and DF21. Bars show the contribution of spectral features ($w_{SF}$, blue) and SSL features ($w_{SSL}$, orange), with solid fill for ITW and hatched fill for DF21, for each feature type (CQCC, LFCC, MFCC).}
    \label{fig:sf_weights}
\end{figure}

Figure~\ref{fig:sf_weights} reports the average normalized weights \(\mathbf{w}_{SF}\) and \(\mathbf{w}_{SSL}\) as shown in Section \ref{sec:fusion}, assigned by the learnable gating module to the spectral and SSL branches, respectively, on the DF21 and ITW evaluation sets. By construction, 
\(\mathbf{w}_{SF}(t) + \mathbf{w}_{SSL}(t) = 1\) for every frame \(t\), so higher \(\mathbf{w}_{SF}\) implies greater reliance on hand‑crafted cues.

Across the three spectral representations, the model consistently allocates approximately 15–22\% of its weight to the spectral branch and 78–85\% to the SSL branch. This results demonstrate that: (i) Although SSL features dominate the fusion, spectral features still account for a stable $\approx 20\%$ of the fused representation. This indicates that the handcrafted features contribute a significant amount of spoof-related information. (ii) The near‑identical weight distributions on DF21 and ITW indicate that our gating mechanism learns a general‑purpose fusion policy, rather than overfitting to dataset‑specific artifacts.


These results validate this hybrid approach for deepfake detection, by blending both views, the model retains the spoofing cues of hand‑crafted while exploiting the rich context captured by SSL features.

\vspace{0.2cm}

\subsection{Comparison with SOTA models}

\begin{table}[]
\centering
\caption{Comparison of EER (\%) on DF21 and ITW for various deepfake detection methods, including our best cross-attention fusion of Wav2Vec2.0 and CQCC.}
\scalebox{1}{
\begin{tabular}{lrrr}
\toprule
                                & \multicolumn{1}{l}{\textbf{DF21}} & \multicolumn{1}{l}{\textbf{ITW}} & \multicolumn{1}{l}{\textbf{AVG}} \\ \hline
\textbf{Wav2Vec2.0 Ensembling \cite{rosello2024anti}}  & 8.74                              & 18.60                             & 13.67                            \\
\textbf{Wav2Vec2.0 AASIST \cite{tak2022automatic}}  & 5.29                              & 14.03                             & 9.66                            \\
\textbf{WavLM ASP+MLP \cite{tran2024spoofed}}          & 4.47                              & 12.87                            & 8.67                             \\
\textbf{SLIM \cite{zhu2024slim}}                   & 4.40                               & 12.50                             & 8.45                             \\
\textbf{MoE \cite{wang2025mixture}}                    & 2.54                              & 9.17                             & 5.86                             \\
\textbf{TCM \cite{truong2024temporal}}                    & 2.06                              & 7.79                             & 4.93                             \\
\textbf{WaveSpec \cite{jin2025wave}}               & 2.01                              & 7.26                             & 4.64                             \\
\textbf{Nes2Net-X \cite{liu2025nes2net}}              & \textbf{1.91}                     & 6.60                              & \textbf{4.26}                    \\ \hline

\hline
\multicolumn{4}{c}{\textit{\textbf{Cross-Attention (SSL \textrightarrow SF)}}} \\ \hline

\hline                                          
          
\textbf{Wav2Vec2.0+CQCC (Ours)} & 2.71                              & \textbf{6.03}                    & 4.37                             \\ \bottomrule
\end{tabular}}
\label{tab:sotacompare}
\end{table}




Table \ref{tab:sotacompare} positions our cross-attention fusion of Wav2Vec2.0 embeddings and CQCC against recent SOTA models on both DF21 and ITW. While Nes2Net-X and WaveSpec attain the lowest EER on DF21, our model matches their DF21 performance closely (2.71\% vs. 1.91\%) and establishes a new best result on the more challenging ITW set compared with all newly reported SOTA models.

Crucially, this competitive edge is achieved without resorting to large ensembles compared to \cite{rosello2024anti}, multi-branch architectures \cite{truong2024temporal}, or dataset-specific augmentations (e.g., Rawboost \cite{tak2022rawboost} used for DF21). Instead, our approach rests on a single, lightweight cross-attention module that dynamically balances handcrafted spectral features and self-supervised representations. This not only simplifies the model but also makes fusion explicitly interpretable (cf.\ Figure \ref{fig:sf_weights}).

In summary, our approach delivers: (i) Strong results on both in-the-wild and controlled datasets with no domain-specific tuning. (ii) Clear gating weights that reveal how much each feature source contributes to the fused model.

These results demonstrate that integrating spectral features with SSL representations enhances complementary information critical for enhancing cross-domain deepfake detection robustness.

\section{Conclusion}
We have analyzed how the fusion framework that combines SSL features with classical spectral features (MFCC, LFCC, CQCC) reinforces deepfake detection under severe domain shifts. By systematically comparing four fusion strategies, concatenation, cross-attention, mutual cross-attention, and learnable gating, within a GNN classifier, we demonstrate consistent and substantial gains over an SSL-only baseline across four benchmarks (LA19, DF21, ITW, ASV5). 
Through learnable gating analysis, we verify that spectral cues contribute roughly 20\% of the fused representation in a dataset-agnostic manner. In our Future work, we will explore these findings with different backend classifiers, and with the incorporation of additional handcrafted descriptors from formant trajectories and prosody features extracted from OpenSmile \cite{eyben2010opensmile} with SSL features as complementary features to enrich the fusion space.




\clearpage
\bibliographystyle{IEEEtran}
\bibliography{refs25}

\begin{thebibliography}{10}
\providecommand{\url}[1]{#1}
\csname url@samestyle\endcsname
\providecommand{\newblock}{\relax}
\providecommand{\bibinfo}[2]{#2}
\providecommand{\BIBentrySTDinterwordspacing}{\spaceskip=0pt\relax}
\providecommand{\BIBentryALTinterwordstretchfactor}{4}
\providecommand{\BIBentryALTinterwordspacing}{\spaceskip=\fontdimen2\font plus
\BIBentryALTinterwordstretchfactor\fontdimen3\font minus \fontdimen4\font\relax}
\providecommand{\BIBforeignlanguage}[2]{{%
\expandafter\ifx\csname l@#1\endcsname\relax
\typeout{** WARNING: IEEEtran.bst: No hyphenation pattern has been}%
\typeout{** loaded for the language `#1'. Using the pattern for}%
\typeout{** the default language instead.}%
\else
\language=\csname l@#1\endcsname
\fi
#2}}
\providecommand{\BIBdecl}{\relax}
\BIBdecl

\bibitem{pham2025comprehensive}
L.~Pham, P.~Lam, D.~Tran, H.~Tang, T.~Nguyen, A.~Schindler, F.~Skopik, A.~Polonsky, and H.~C. Vu, ``A comprehensive survey with critical analysis for deepfake speech detection,'' \emph{Computer Science Review}, vol.~57, p. 100757, 2025.

\bibitem{hao2025boosting}
H.~Hao, L.~Zhou, S.~Liu, J.~Li, S.~Hu, R.~Wang, and F.~Wei, ``Boosting large language model for speech synthesis: An empirical study,'' in \emph{ICASSP 2025-2025 IEEE International Conference on Acoustics, Speech and Signal Processing (ICASSP)}.\hskip 1em plus 0.5em minus 0.4em\relax IEEE, 2025, pp. 1--5.

\bibitem{zhang2023speechgpt}
D.~Zhang, S.~Li, X.~Zhang, J.~Zhan, P.~Wang, Y.~Zhou, and X.~Qiu, ``Speechgpt: Empowering large language models with intrinsic cross-modal conversational abilities,'' \emph{arXiv preprint arXiv:2305.11000}, 2023.

\bibitem{rubenstein2023audiopalm}
P.~K. Rubenstein, C.~Asawaroengchai, D.~D. Nguyen, A.~Bapna, Z.~Borsos, F.~d.~C. Quitry, P.~Chen, D.~E. Badawy, W.~Han, E.~Kharitonov \emph{et~al.}, ``Audiopalm: A large language model that can speak and listen,'' \emph{arXiv preprint arXiv:2306.12925}, 2023.

\bibitem{meaker2023deepfake}
M.~Meaker, ``Deepfake audio is a political nightmare,'' 2023.

\bibitem{mcafee2023beware}
McAfee, ``Beware the artificial impostor: A mcafee cybersecurity artificial intelligence report\_2023,'' May 2023.

\bibitem{chen2021ur}
X.~Chen, Y.~Zhang, G.~Zhu, and Z.~Duan, ``Ur channel-robust synthetic speech detection system for asvspoof 2021,'' \emph{arXiv preprint arXiv:2107.12018}, 2021.

\bibitem{caceres2021biometric}
J.~C{\'a}ceres, R.~Font, T.~Grau, J.~Molina, and B.~V. SL, ``The biometric vox system for the asvspoof 2021 challenge,'' in \emph{Proc. ASVspoof2021 Workshop}, 2021.

\bibitem{das2021known}
R.~K. Das, ``Known-unknown data augmentation strategies for detection of logical access, physical access and speech deepfake attacks: Asvspoof 2021,'' \emph{Proc. 2021 Edition of the Automatic Speaker Verification and Spoofing Countermeasures Challenge}, pp. 29--36, 2021.

\bibitem{tomilov2021stc}
A.~Tomilov and A.~e.~a. Svishchev, ``Stc antispoofing systems for the asvspoof2021 challenge,'' in \emph{Proc. ASVspoof 2021 Workshop}, 2021, pp. 61--67.

\bibitem{muller2022does}
N.~M. M{\"u}ller, P.~Czempin, F.~Dieckmann, A.~Froghyar, and K.~B{\"o}ttinger, ``Does audio deepfake detection generalize?'' \emph{arXiv preprint arXiv:2203.16263}, 2022.

\bibitem{alenin2022subnetwork}
A.~Alenin, N.~Torgashov, A.~Okhotnikov, R.~Makarov, and I.~Yakovlev, ``A subnetwork approach for spoofing aware speaker verification.'' in \emph{INTERSPEECH}, 2022, pp. 2888--2892.

\bibitem{chen2021pindrop}
T.~Chen, E.~Khoury, K.~Phatak, and G.~Sivaraman, ``Pindrop labs’ submission to the asvspoof 2021 challenge,'' \emph{Proc. 2021 edition of the automatic speaker verification and spoofing countermeasures challenge}, pp. 89--93, 2021.

\bibitem{dong2023multi}
S.~Dong, J.~Xue, C.~Fan, K.~Zhu, Y.~Chen, and Z.~Lv, ``Multi-perspective information fusion res2net with randomspecmix for fake speech detection,'' \emph{arXiv preprint arXiv:2306.15389}, 2023.

\bibitem{wang2021comparative}
X.~Wang and J.~Yamagishi, ``A comparative study on recent neural spoofing countermeasures for synthetic speech detection,'' \emph{arXiv preprint arXiv:2103.11326}, 2021.

\bibitem{wang2024multi}
C.~Wang, J.~He, J.~Yi, J.~Tao, C.~Y. Zhang, and X.~Zhang, ``Multi-scale permutation entropy for audio deepfake detection,'' in \emph{ICASSP 2024-2024 IEEE International Conference on Acoustics, Speech and Signal Processing (ICASSP)}.\hskip 1em plus 0.5em minus 0.4em\relax IEEE, 2024, pp. 1406--1410.

\bibitem{ge2021raw}
W.~Ge, J.~Patino, M.~Todisco, and N.~Evans, ``Raw differentiable architecture search for speech deepfake and spoofing detection,'' \emph{arXiv preprint arXiv:2107.12212}, 2021.

\bibitem{jung2022aasist}
J.-w. Jung, H.-S. Heo, H.~Tak, H.-j. Shim, J.~S. Chung, B.-J. Lee, H.-J. Yu, and N.~Evans, ``Aasist: Audio anti-spoofing using integrated spectro-temporal graph attention networks,'' in \emph{ICASSP 2022-2022 IEEE}.\hskip 1em plus 0.5em minus 0.4em\relax IEEE, 2022, pp. 6367--6371.

\bibitem{xie2023learning}
Y.~Xie, H.~Cheng, Y.~Wang, and L.~Ye, ``Learning a self-supervised domain-invariant feature representation for generalized audio deepfake detection,'' in \emph{Proc. Interspeech}, vol. 2023, no. 2023, 2023, pp. 2808--2812.

\bibitem{wang2024can}
X.~Wang and J.~Yamagishi, ``Can large-scale vocoded spoofed data improve speech spoofing countermeasure with a self-supervised front end?'' in \emph{ICASSP 2024-2024 IEEE International Conference on Acoustics, Speech and Signal Processing (ICASSP)}.\hskip 1em plus 0.5em minus 0.4em\relax IEEE, 2024, pp. 10\,311--10\,315.

\bibitem{yang2024robust}
Y.~Yang, H.~Qin, H.~Zhou, C.~Wang, T.~Guo, K.~Han, and Y.~Wang, ``A robust audio deepfake detection system via multi-view feature,'' in \emph{ICASSP 2024-2024 IEEE International Conference on Acoustics, Speech and Signal Processing}.\hskip 1em plus 0.5em minus 0.4em\relax IEEE, 2024, pp. 13\,131--13\,135.

\bibitem{muller2024harder}
N.~M. M{\"u}ller, N.~Evans, H.~Tak, P.~Sperl, and K.~B{\"o}ttinger, ``Harder or different? understanding generalization of audio deepfake detection,'' \emph{arXiv preprint arXiv:2406.03512}, 2024.

\bibitem{berrebbi2022combining}
D.~Berrebbi, J.~Shi, B.~Yan, O.~L{\'o}pez-Francisco, J.~D. Amith, and S.~Watanabe, ``Combining spectral and self-supervised features for low resource speech recognition and translation,'' \emph{arXiv preprint arXiv:2204.02470}, 2022.

\bibitem{peng2024fine}
S.~Peng, W.~Guo, H.~Wu, Z.~Li, and J.~Zhang, ``Fine-tune pre-trained models with multi-level feature fusion for speaker verification,'' in \emph{Proc. Interspeech}, 2024, pp. 2110--2114.

\bibitem{jin2025wave}
Z.~Jin, L.~Lang, and B.~Leng, ``Wave-spectrogram cross-modal aggregation for audio deepfake detection,'' in \emph{ICASSP 2025-2025 IEEE International Conference on Acoustics, Speech and Signal Processing (ICASSP)}.\hskip 1em plus 0.5em minus 0.4em\relax IEEE, 2025, pp. 1--5.

\bibitem{zhang2025audio}
B.~Zhang, H.~Cui, V.~Nguyen, and M.~Whitty, ``Audio deepfake detection: What has been achieved and what lies ahead,'' \emph{Sensors (Basel, Switzerland)}, vol.~25, no.~7, p. 1989, 2025.

\bibitem{kheir2025comprehensive}
Y.~E. e.~a. Kheir, ``Comprehensive layer-wise analysis of ssl models for audio deepfake detection,'' \emph{arXiv preprint arXiv:2502.03559}, 2025.

\bibitem{wang2020asvspoof}
X.~Wang, J.~Yamagishi, M.~Todisco, H.~Delgado, A.~Nautsch, N.~Evans, M.~Sahidullah, V.~Vestman, T.~Kinnunen, K.~A. Lee \emph{et~al.}, ``Asvspoof 2019: A large-scale public database of synthesized, converted and replayed speech,'' \emph{Computer Speech \& Language}, vol.~64, p. 101114, 2020.

\bibitem{yamagishi2021asvspoof}
J.~Yamagishi, X.~Wang, M.~Todisco, M.~Sahidullah, J.~Patino, A.~Nautsch, X.~Liu, K.~A. Lee, T.~Kinnunen, N.~Evans \emph{et~al.}, ``Asvspoof 2021: accelerating progress in spoofed and deepfake speech detection,'' in \emph{ASVspoof 2021 Workshop-Automatic Speaker Verification and Spoofing Coutermeasures Challenge}, 2021.

\bibitem{wang2024asvspoof5}
\BIBentryALTinterwordspacing
X.~Wang, H.~Delgado, H.~Tak, J.~weon Jung, H.~jin Shim, M.~Todisco, I.~Kukanov, X.~Liu, M.~Sahidullah, T.~Kinnunen, N.~Evans, K.~A. Lee, and J.~Yamagishi, ``Asvspoof 5: Crowdsourced speech data, deepfakes, and adversarial attacks at scale,'' 2024. [Online]. Available: \url{https://arxiv.org/abs/2408.08739}
\BIBentrySTDinterwordspacing

\bibitem{diederik2014adam}
\BIBentryALTinterwordspacing
D.~P. Kingma and J.~Ba, ``Adam: A method for stochastic optimization,'' 2017. [Online]. Available: \url{https://arxiv.org/abs/1412.6980}
\BIBentrySTDinterwordspacing

\bibitem{laakkonen2025generalizable}
J.~Laakkonen, I.~Kukanov, and V.~Hautam{\"a}ki, ``Generalizable speech deepfake detection via meta-learned lora,'' \emph{arXiv preprint arXiv:2502.10838}, 2025.

\bibitem{todisco2017constant}
M.~Todisco, H.~Delgado, and N.~Evans, ``Constant q cepstral coefficients: A spoofing countermeasure for automatic speaker verification,'' \emph{Computer Speech \& Language}, vol.~45, pp. 516--535, 2017.

\bibitem{rosello2024anti}
E.~Rosell{\'o}~Casado, {\'A}.~M. G{\'o}mez~Garc{\'\i}a, I.~L{\'o}pez~Espejo, A.~M. Peinado~Herreros, J.~M. Mart{\'\i}n~Do{\~n}as \emph{et~al.}, ``Anti-spoofing ensembling model: Dynamic weight allocation in ensemble models for improved voice biometrics security,'' 2024.

\bibitem{tak2022automatic}
H.~Tak, M.~Todisco, X.~Wang, J.-w. Jung, J.~Yamagishi, and N.~Evans, ``Automatic speaker verification spoofing and deepfake detection using wav2vec 2.0 and data augmentation,'' \emph{arXiv preprint arXiv:2202.12233}, 2022.

\bibitem{tran2024spoofed}
H.~M. Tran, D.~Guennec, P.~Martin, A.~Sini, D.~Lolive, A.~Delhay, and P.-F. Marteau, ``Spoofed speech detection with a focus on speaker embedding,'' in \emph{INTERSPEECH 2024}, 2024.

\bibitem{zhu2024slim}
Y.~Zhu, S.~Koppisetti, T.~Tran, and G.~Bharaj, ``Slim: Style-linguistics mismatch model for generalized audio deepfake detection,'' \emph{Advances in Neural Information Processing Systems}, vol.~37, pp. 67\,901--67\,928, 2024.

\bibitem{wang2025mixture}
Z.~Wang, R.~Fu, Z.~Wen, J.~Tao, X.~Wang, Y.~Xie, X.~Qi, S.~Shi, Y.~Lu, Y.~Liu \emph{et~al.}, ``Mixture of experts fusion for fake audio detection using frozen wav2vec 2.0,'' in \emph{ICASSP 2025-2025 IEEE International Conference on Acoustics, Speech and Signal Processing (ICASSP)}.\hskip 1em plus 0.5em minus 0.4em\relax IEEE, 2025, pp. 1--5.

\bibitem{truong2024temporal}
D.-T. Truong, R.~Tao, T.~Nguyen, H.-T. Luong, K.~A. Lee, and E.~S. Chng, ``Temporal-channel modeling in multi-head self-attention for synthetic speech detection,'' \emph{arXiv preprint arXiv:2406.17376}, 2024.

\bibitem{liu2025nes2net}
T.~Liu, D.-T. Truong, R.~K. Das, K.~A. Lee, and H.~Li, ``Nes2net: A lightweight nested architecture for foundation model driven speech anti-spoofing,'' \emph{arXiv preprint arXiv:2504.05657}, 2025.

\bibitem{tak2022rawboost}
H.~Tak, M.~Kamble, J.~Patino, M.~Todisco, and N.~Evans, ``Rawboost: A raw data boosting and augmentation method applied to automatic speaker verification anti-spoofing,'' in \emph{ICASSP 2022-2022 IEEE International Conference on Acoustics, Speech and Signal Processing (ICASSP)}.\hskip 1em plus 0.5em minus 0.4em\relax IEEE, 2022, pp. 6382--6386.

\bibitem{eyben2010opensmile}
F.~Eyben, M.~W{\"o}llmer, and B.~Schuller, ``Opensmile: the munich versatile and fast open-source audio feature extractor,'' in \emph{Proceedings of the 18th ACM international conference on Multimedia}, 2010, pp. 1459--1462.

\end{thebibliography}

\end{document}